\documentclass{an}
\usepackage{graphicx}
\usepackage{times}
\usepackage{fancyhdr}
\newcommand{\EQ}{\begin{equation}}
\newcommand{\EN}{\end{equation}}
\newcommand{\EQA}{\begin{eqnarray}}
\newcommand{\ENA}{\end{eqnarray}}
\newcommand{\eq}[1]{(\ref{#1})}
\newcommand{\EEq}[1]{Equation~(\ref{#1})}
\newcommand{\Eq}[1]{Eq.~(\ref{#1})}

\newcommand{\Sec}[1]{Sect.~\ref{#1}}

\newcommand{\Fig}[1]{Fig.~\ref{#1}}

\newcommand{\bra}[1]{\langle #1\rangle}

\newcommand{\meanF}{\overline{\cal F}} 
\newcommand{\meanEMF}{\overline{\vec{\cal E}}}
\newcommand{\meanemf}{\overline{\cal E}}
\newcommand{\meanFF}{\overline{\mbox{\boldmath ${\cal F}$}} {}}

\newcommand{\meanB}{\overline{B}}
\newcommand{\meanJ}{\overline{J}}
\newcommand{\meanU}{\overline{U}}
\newcommand{\meanC}{\overline{C}}

\newcommand{\meanBB}{\overline{\vec{B}}}
\newcommand{\meanJJ}{\overline{\vec{J}}}
\newcommand{\meanUU}{\overline{\vec{U}}}
\newcommand{\meanWW}{\overline{\vec{W}}}

\newcommand{\meanEE}{\overline{\vec{E}}}
\newcommand{\xx}{\mbox{\boldmath $x$} {}}

\newcommand{\ddelta}{{\vec{\delta}}}
\newcommand{\ggamma}{{\vec{\gamma}}}

\newcommand{\uu}{{\vec{u}}}
\newcommand{\BB}{{\vec{B}}}
\newcommand{\JJ}{{\vec{J}}}
\newcommand{\jj}{{\vec{j}}}
\newcommand{\AAA}{{\vec{A}}}
\newcommand{\aaaa}{{\vec{a}}}
\newcommand{\bb}{{\vec{b}}}
\newcommand{\cc}{{\vec{c}}}

\newcommand{\ee}{\mbox{\boldmath $e$} {}}

\newcommand{\EE}{{\vec{E}}}

\newcommand{\TT}{\mbox{\boldmath $T$} {}}

\newcommand{\nab}{\mbox{\boldmath $\nabla$} {}}

\newcommand{\oo}{\mbox{\boldmath $\omega$} {}}

\newcommand{\emf}{\mbox{\boldmath ${\cal E}$} {}}

\newcommand{\DD}{{\rm D} {}}
\newcommand{\dd}{{\rm d} {}}
\newcommand{\const}{{\rm const}  {}}

\def\Pra{\mbox{\rm Pr}}

\def\half{{\textstyle{1\over2}}}

\def\onethird{{\textstyle{1\over3}}}

\newcommand{\yjgr}[3]{ #1, {JGR,} {#2}, #3}

\newcommand{\yapj}[3]{ #1, {ApJ,} {#2}, #3}
\newcommand{\yapjl}[3]{ #1, {ApJ,} {#2}, #3}

\newcommand{\yan}[3]{ #1, {AN,} {#2}, #3}

\newcommand{\yana}[3]{ #1, {A\&A,} {#2}, #3}

\newcommand{\ygafd}[3]{ #1, {Geophys. Astrophys. Fluid Dyn.,} {#2}, #3}
\newcommand{\yjfm}[3]{ #1, {JFM,} {#2}, #3}
\newcommand{\ypf}[3]{ #1, {Phys. Fluids,} {#2}, #3}

\newcommand{\yjetp}[3]{ #1, {Sov. Phys. JETP,} {#2}, #3}

\newcommand{\yprl}[3]{ #1, {PRL,} {#2}, #3}
\newcommand{\ypre}[3]{ #1, {PRE,} {#2}, #3}

\newcommand{\yptrs}[3]{ #1, {Phil. Trans. Roy. Soc.,} {#2}, #3}
\newcommand{\ymn}[3]{ #1, {MNRAS,} {#2}, #3}

\newcommand{\ysph}[3]{ #1, {Solar Phys.,} {#2}, #3}

\newcommand{\yjour}[4]{ #1, {#2}, {#3}, #4}

\newcommand{\ybook}[3]{ #1, {#2} (#3)}
\newcommand{\yproc}[5]{ #1, in {#3}, ed. #4 (#5), #2}

\newcommand{\tapj}[1]{ #1, {ApJ,} (to be submitted)}

\newcommand{\pan}[1]{ #1, {AN,} (in press)}

\newcommand{\sjour}[2]{ #1, {#2,} (submitted)}
\begin{document}
\setcounter{page}{1}

\title{The problem of small and large scale fields in the solar dynamo}

\author{A.\ Brandenburg$^{1,2}$, N.\ E.\ L.\ Haugen$^{3,4}$,
P.\ J.\ K\"apyl\"a$^{5,6}$ and C.\ Sandin$^7$}
\institute{
$^1$Isaac Newton Institute for Mathematical Sciences,
20 Clarkson Road, Cambridge CB3 0EH, UK\\
$^2$Nordita, Blegdamsvej 17, DK-2100 Copenhagen \O, Denmark\\
$^3$DAMTP, University of Cambridge,
Wilberforce Road, Cambridge CB3 0WA, UK\\
$^4$Dept.\ of Physics, The Norwegian University of Science and Technology, 
H{\o}yskoleringen 5, N-7034 Trondheim, Norway\\
$^5$Kiepenheuer-Institut f\"ur Sonnenphysik, Sch\"oneckstra{\ss}e 6,
D-79104 Freiburg, Germany \\
$^6$Dept.\ of Physical Sciences, Astronomy Div., P.O. Box 3000,
FIN-90014 University of Oulu, Finland\\
$^7$Stockholm, Sweden
}

\date{Received; accepted; published online}

\abstract{
Three closely related stumbling blocks of solar mean field dynamo theory
are discussed: how dominant are the small scale fields, how is the alpha
effect quenched, and whether magnetic and current helicity fluxes alleviate
the quenching?
It is shown that even at the largest currently available resolution
there is no clear evidence of power law scaling of the magnetic and kinetic
energy spectra in turbulence.
However, using subgrid scale modeling, some indications of asymptotic
equipartition can be found.
The frequently used first order smoothing approach to calculate
the alpha effect and other transport coefficients is contrasted with
the superior minimal tau approximation.
The possibility of catastrophic alpha quenching is discussed as a result
of magnetic helicity conservation.
Magnetic and current helicity fluxes are shown to alleviate catastrophic
quenching in the presence of shear.
Evidence for strong large scale dynamo action, even in the absence of
helicity in the forcing, is presented.
}

\correspondence{brandenb@nordita.dk}

\maketitle

\section{Introduction}

Over the past 30 years, the standard approach to understanding the origin
of the solar cycle has been mean field dynamo theory.
This approach can be justified simply by the fact that the sun does have
a finite azimuthally averaged mean field, $\meanBB(r,\theta,t)$, where
$r$ and $\theta$ are radius and colatitude.
Observationally, we really only know with some certainty its radial
component at the surface, $\meanB_r(R,\theta,t)$, where $R$ is the radius
of the sun.
There is also some indirect evidence for the toroidal field,
$\meanB_\phi(r_{\rm spot},\theta,t)$, where $r_{\rm spot}$ is the not
well known radius where sunspots are anchored.
Both components give a very clear indication of the spatio-temporal coherence
of the mean field, with a 22 year cycle and latitudinal migration.

Mean field theory has certainly been successful in showing that a
solar-like mean field can be produced if there is an $\alpha$ effect,
i.e.\ if the mean electromotive force has a component along the mean field
(Weiss 2005).
A number of complications have arisen in the mean time.
\begin{itemize}
\item[(i)] The standard theory for calculating the value of $\alpha$
and other relevant transport coefficients (such as turbulent magnetic
diffusivity) relies on the first order smoothing (FOSA) or second order
correlation approximation (SOCA).
This approach is valid if either the magnetic Reynolds number is small
(poor microscopic conductivity of the gas) or if the so-called Strouhal
number (K\"apyl\"a et al.\ 2005) is small compared to unity.
The latter means that the correlation time is supposed to be much less
than the turnover time of the turbulence, which is not usually the case.
The former assumption is equally inappropriate.
\item[(ii)] Shortly after mean field theory became popular there have
been recurrent concerns about its applicability when the field strength
is comparable to the equipartition value of the turbulence, i.e.\ if the
magnetic energy is comparable to the kinetic energy of the turbulence.
Such concerns where first addressed by Piddington (1970, 1972), but more
recently by Vainshtein \& Cattaneo (1992) and Kulsrud \& Anderson (1993).
\item[(iii)] In more recent years the problem of the resistively slow
evolution of magnetic helicity has been discussed in connection with
a correspondingly slow saturation if the dynamo (Brandenburg 2001,
Mininni et al.\ 2003).
This resistively slow time scale in the problem can also affect the
cycle period in an $\alpha\Omega$ dynamo (Brandenburg et al.\ 2001, 2002).
The presence of boundaries may help (Blackman \& Field 2000a,b, Kleeorin
et al.\ 2000, 2002, 2003), but it may also make matters worse
(Brandenburg \& Dobler 2001).
The question is therefore whether suitably arranged shear is needed
to transport magnetic helicity out of the domain (Vishniac \& Cho 2001,
Subramanian \& Brandenburg 2004).
The indications are that with shear and open boundary conditions
the otherwise cata\-strophic $\alpha$ quenching can be alleviated
(Brandenburg \& Sandin 2004).
\end{itemize}

In the following we discuss the current status on all three issues.
Indeed, there has been a lot of progress, and from an optimistic viewpoint
one might almost think that these issues have now been solved.
But, of course, as always in science, new problems and unexpected issues
emerge all the time.
Also, having proposed one solution to a problem does not exclude alternative
solutions, and so only with time will we be able to look back and say how
it really was.

\section{Problem~I: dominance of small scale fields?}

The first issue of the relative importance of small scale versus large
scale fields can be discussed in terms of the magnetic energy spectrum.
We begin with a historical perspective.

\subsection{Turbulent diffusion and turbulent cascade}

\begin{figure}[t!]\centering\includegraphics[width=\columnwidth]
{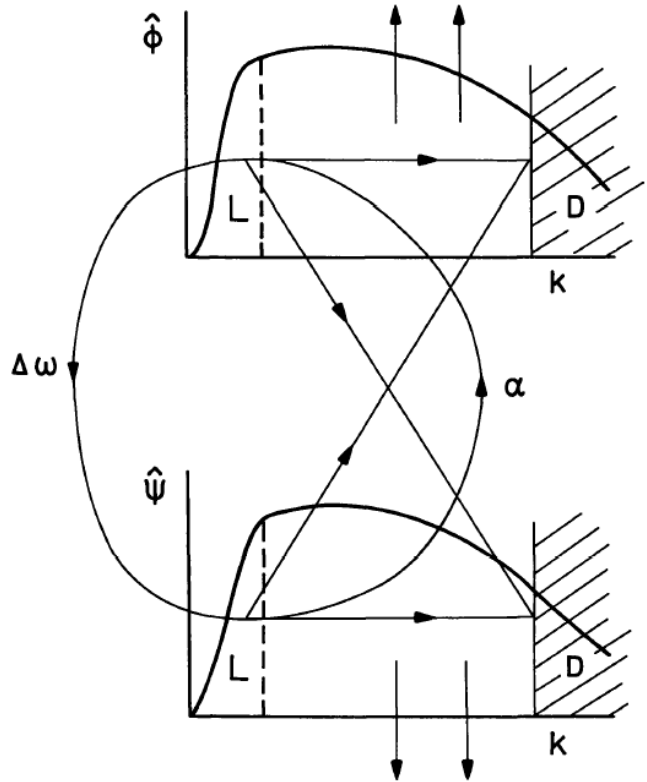}\caption{
A sketch from the original review by Stix (1974) where he illustrates
the role of a turbulent cascade in converting large scale poloidal and
toroidal flux into small scale fields where it can be converted into
heat by microscopic resistivity.
}\label{stix74c}\end{figure}

Turbulent diffusion relies on the ability of the turbulence to transport
energy from large scales to small scales.
This is an important consideration that was discussed by Stix (1974)
in the context of early criticism raised against dynamo theory.
The situation is illustrated in \Fig{stix74c}.

The assumption of a magnetic energy spectrum being parallel or even coincident
with that of kinetic energy, and this being equivalent to the cascade
in ordinary (nonmagnetic) turbulence is natural, but not trivial and not
fully confirmed by simulations, as will be discussed in the next section.

\subsection{Numerical indications for a magnetic cascade}

\begin{figure*}[t!]\centering\includegraphics[width=\textwidth]
{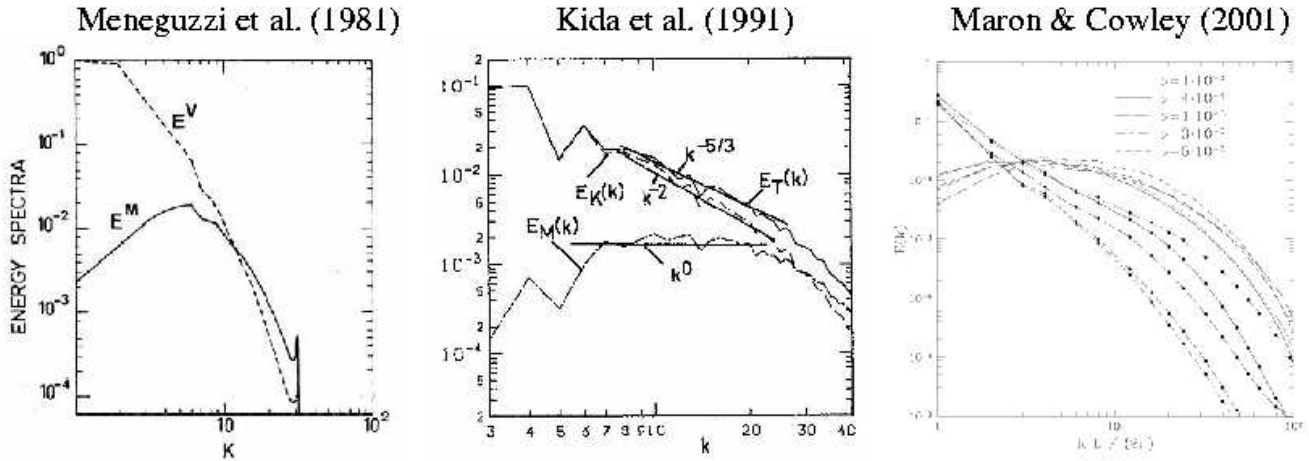}\caption{
Steady progress in solving the small scale dynamo problem.
Kinetic and magnetic energy spectra from the papers by
Meneguzzi et al.\ (1981), Kida et al.\ (1991), and Maron \& Cowley (2001).
The numerical resolution used in the three cases is $64^3$, $128^3$, and
$256^3$ meshpoints, respectively.
All spectra agree in that at large scales (small wavenumbers $k$) the
spectral magnetic energy is below the spectral kinetic energy.
In the spectra of Meneguzzi et al.\ (1981) and Maron \& Cowley (2001)
the spectral magnetic energy peaks at about $k=5$.
}\label{menekidamaron}\end{figure*}

\begin{figure}[t!]\centering\includegraphics[width=\columnwidth]
{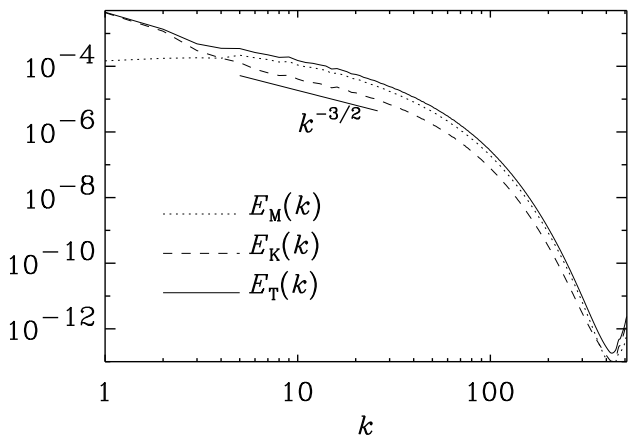}\caption{
Magnetic, kinetic and total energy spectra.
$1024^3$ meshpoints.
The Reynolds number is $u_{\rm rms}/(\nu k_{\rm f})\approx960$.
[Adapted from Haugen et al.\ (2003).]
}\label{power1024a}\end{figure}

Over the past decades some steady progress has been brought about;
see \Fig{menekidamaron}, where we show kinetic and magnetic energy
spectra from Meneguzzi et al.\ (1981), Kida et al.\ (1991),
Maron \& Cowley (2001), representing the improvement of the state
of the art simulations over the past three decades.
The spectra from all three papers
show that at large scales the spectral magnetic energy
is below the spectral kinetic energy.
In the spectra of Meneguzzi et al.\ (1981) and Maron \& Cowley (2001)
the spectral magnetic energy peaks at about $k\approx5$.
However, there has not really been any evidence for power law behavior.
One exception is the spectrum obtained by Kida et al.\ (1991) who find
a $k^0$ spectrum, but this has not been found in subsequent simulations by
Maron \& Cowley (2001) or Maron et al.\ (2004), for example.
The numerical resolution used in the three cases is $64^3$, $128^3$, and
$256^3$ meshpoints, respectively.
Obviously, much higher resolution is needed to begin to address the
possibility of powerlaw behavior.
This was the main reason for Haugen et al.\ (2003) to push the resolution
to $1024^3$ meshpoints; see \Fig{power1024a}.
Based on these results, one can begin to see the development of what
looks like an inertial range with a tentative $k^{-3/2}$ scaling.
If this is confirmed, it would rule out earlier claims that the magnetic
energy spectrum peaks at the resistive scale (Maron \& Blackman 2002,
Schekochihin et al.\ 2002).
However, the results by Haugen et al.\ (2003) are still open to
alternative interpretations.
Schekochihin et al.\ (2004a) have argued that the magnetic spectrum
is still curved, and that there is therefore actually no evidence for
powerlaw behavior.
So, the asymptotic spectral behavior of hydromagnetic turbulence is
still very much an open question.

A puzzling aspect of all the magnetic energy spectra is the excess
of spectral magnetic energy over spectral kinetic energy.
In order to get some insight into the possible asymptotic behavior,
Haugen \& Brandenburg (2004) have recently considered simulations
with subgrid scale modeling using either hyperviscosity or Smagorinsky
viscosity together with magnetic hyper-resistivity.

\begin{figure}\centering\includegraphics[width=\columnwidth]{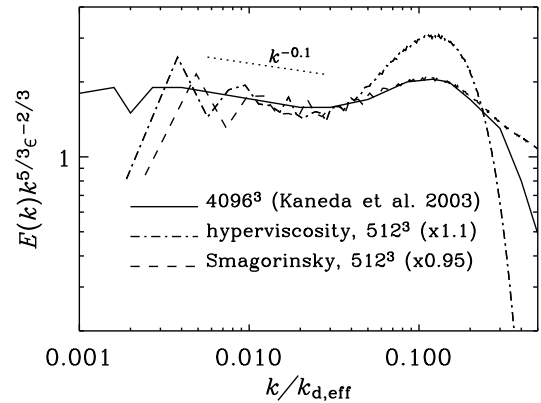}
\caption{Comparison of energy spectra of the $4096^3$ mesh points run
of Kaneda et al.\ (solid line) and $512^3$ mesh points runs with
hyperviscosity (dash-dotted line) and Smagorinsky viscosity (dashed line).}
\label{kan_hyp_smag}
\end{figure}

Consider for comparison first the purely hydrodynamic case
(\Fig{kan_hyp_smag}), where we can use the high resolution simulations
of Kaneda et al.\ (2003) on the {\sc Earth Simulator} as benchmark.
This simulation corresponds to a resolution of $4096$ collocation points.
Note that there is not even a clear confirmation of the famous
Kolmogorov $k^{-5/3}$ spectrum, but there is a $k^{-0.1}$ correction
in what looks like the inertial range.
However, simulations with both hyperviscosity (Haugen \& Brandenburg 2004)
and Smagorinsky subgrid scale modeling
(Haugen \& Brandenburg, in preparation) confirm
this correction to the inertial range; see \Fig{kan_hyp_smag}.

The uprise in the compensated power spectra just before the dissipative
subrange is due to the bottleneck effect in turbulence (Falkovich 1994).
This effect is much weaker in wind tunnel turbulence (She \& Jackson 1993),
but this is because these are one-dimensional spectra, $E_{\rm 1D}(k)$,
which are related to the fully three-dimensional spectra via a simple
integral transformation
\EQ
E_{\rm 1D}(k)=\int_k^\infty{E_{\rm 3D}(k')\over k'}\,\dd k'.
\EN
Of course, for perfect power law spectra the two are the same, but they
can be quite different when there are departures from power law behavior,
such as due to the bottleneck effect itself and due to the dissipative subrange
(Dobler et al.\ 2003).

\begin{figure}\centering\includegraphics[width=0.5\textwidth]
{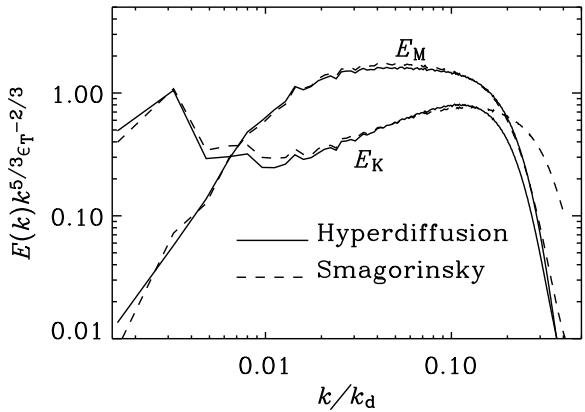}\caption{
Magnetic and kinetic energy spectra for runs with $512^3$
mesh points and hyperviscosity and hyper-resistivity (solid line) and
Smagorinsky viscosity and hyper-resistivity (dashed line).
Note the mutual approach of kinetic and magnetic energy spectra
before entering the dissipative subrange.
}\label{hyper_512}\end{figure}

\begin{figure}\centering\includegraphics[width=0.45\textwidth]
{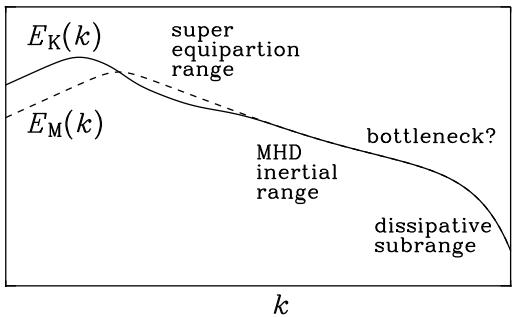}\caption{
Sketch of the anticipated kinetic and magnetic energy
spectra in the large Reynolds number limit for $\Pra_{M}=1$.
Note the slight super-equipartition just to the right
of the peak of $E_{\rm M}(k)$ and the asymptotic 
equipartition for large wavenumbers.
}\label{sketch}\end{figure}

In \Fig{hyper_512} we show the results for the hydromagnetic case where
we have used hyperresistivity and either hyperviscosity or Smagorinsky
subgrid scale modeling for the velocity field.
There are two important things to notice.
First, the compensated magnetic spectrum seems flat, but the kinetic
energy spectrum seems to rise, possibly approaching the magnetic
energy spectrum.
A sketch of what we believe the asymptotic magnetic and kinetic energy
spectra could look like is shown in \Fig{sketch}.

In summary, there seems now some evidence suggesting that there
is indeed a magnetic energy cascade.
We should emphasize, however, that most of the simulations to date
have either unit magnetic Prandtl number, or at least magnetic Prandtl numbers
that are not very different from unity.

\begin{figure}\centering\includegraphics[width=\columnwidth]
{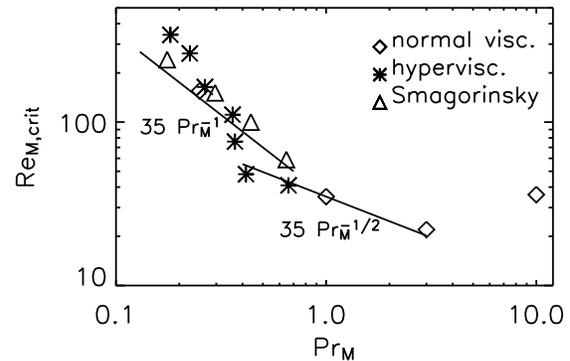}\caption{
Dependence of the critical magnetic Reynolds numbers as a function
of the magnetic Prandtl number using for the direct simulations
(Haugen et al.\ 2004), indicated by diamonds, compared with results
using hyperviscosity (asterisks) or Smagorinsky viscosity (triangles).
}\label{Rmcrit}\end{figure}

The question of the magnetic Prandtl number dependence is potentially
quite important for stars where this number is very small (magnetic
diffusivity much larger than the kinematic viscosity).
Using a modified Kazantsev model, Rogachevskii \& Kleeorin (1997)
pointed out that when the velocity field displays Kolmogorov scaling
near the resistive cutoff wave number, the dynamo becomes much harder
to excite than for a velocity field that shows significant power only
at large scales.
They found that the critical magnetic Reynolds number for dynamo action
can exceed a value of 400, which is about 10 times larger than the
critical value of 35 for unit magnetic Prandtl number.
A similar result has been found more recently by 
Schekochihin et al.\ (2004b) and Boldyrev \& Cattaneo (2004).
Direct simulations suggest that the critical magnetic Reynolds
number increases with decreasing magnetic Prandtl number, at least like
$R_{\rm m,crit}\approx35\,\mbox{Pr}_{\rm M}^{-1/2}$ (Haugen et al.\ 2004);
this result is not really confirmed by simulations with Smagorinsky
viscosity and hyperviscosity which seem to suggest a sharper
dependence (\Fig{Rmcrit}); see Schekochihin et al.\ (2004c).
One might have thought that this could be connected with
the artificially enhanced bottleneck effect in the hyperviscous
simulations, but the idea is not supported by the direct inspection
of the kinetic energy spectra that do not show a bottleneck effect.

In simulations with helicity, that will be discussed in a later
section, the value of the magnetic Prandtl number does not seem to
affect the onset of dynamo action.  Indeed, simulations with a
magnetic Prandtl number of 0.1 (Brandenburg 2001)
showed that the critical value of the
magnetic Reynolds number was the same as for unit magnetic Prandtl
number.

\section{Problem~II: is first order smoothing OK?}

Let us now turn to the question of how to generate large scale fields.
This question can be addressed in terms of mean field dynamo theory
where the $\alpha$ effect (and perhaps other effects) play a central role.
The value of $\alpha$ is traditionally calculated using the first order
smoothing approximation, where one neglects triple correlations.
This approach is in principle not applicable when the magnetic Reynolds
number is large, but it seems to work anyway.
Why this is so, and how one can do better, can perhaps best be understood
in the simpler case of passive scalar diffusion.

\subsection{Passive scalar diffusion}

In this section we describe both the first order smoothing approximation
and the minimal tau approximation.
We follow here the presentation of Blackman \& Field (2003), who studied
the passive scalar case as a simpler test case of the more interesting
magnetic case which they studied earlier (Blackman \& Field 2002).
The minimal tau approximation was first used by
Vainshtein \& Kitchatinov (1986) and Kleeorin et al.\ (1990).

The evolution equation for the passive scalar concentration $C$ is,
in the absence of microscopic diffusion,
\EQ
{\DD C\over\DD t}=0
\label{dCdt}
\EN
where $\DD/\DD t=\partial/\partial t+\uu\cdot\nab$ is the advective
derivative.
We assume, for simplicity, that the flow is incompressible, i.e.\
$\nab\cdot\uu=0$, and that $C$ shows some large scale variation so
that a meaningful average can be defined (denoted by an overbar).
Thus, after averaging we have
\EQ
{\partial\meanC\over\partial t}=-{\partial\over\partial x_j}
\overline{u_j c}.
\label{dCbardt}
\EN
One is obviously interested in a closed equation for the mean
flux of passive scalar concentration,
\EQ
\meanFF\equiv\overline{\uu c},
\EN
in terms of the mean concentration, i.e.\ $\meanFF=\meanFF(\meanC)$.
For the reader who is more familiar with the magnetic case we note that
one should think of $\meanFF$ being similar to the electromotive force
$\meanEMF=\overline{\uu\times\bb}$ where, in turn, one is interested
in a closed equation of the form $\meanEMF=\meanEMF(\meanBB)$.

Subtracting \Eq{dCbardt} from \Eq{dCdt} we obtain the equation for the
fluctuation $c=C-\meanC$,
\EQ
{\partial c\over\partial t}=-\partial_j(u_j\meanC)
-\partial_j(u_j c)-\partial_j\overline{u_j c}.
\EN
This is where we come to a turning point.
In the first order smoothing approximation (FOSA) one calculates
\EQ
\meanFF(t)=\overline{\uu(t)\textstyle{\int}\dot{c}(t')\,\dd t'}
\quad\mbox{(FOSA)},
\EN
where $\dot{c}=\partial c/\partial t$, and we have omitted the common
$\xx$ dependence on all quantities.
In the minimal tau approximation (MTA) one calculates instead
\EQ
{\partial\meanFF\over\partial t}
=\overline{\uu\dot{c}}+\overline{\dot{\uu}c}
\quad\mbox{(MTA)}.
\EN
A principle difference between the two approaches is that the
momentum equation is naturally incorporated under MTA, i.e.\
\EQ
\dot{\uu}=-u_j\partial_j u_i-\partial_i p,
\EN
where $p$ is the pressure.
A more interesting situation would arise if we allowed here
for rotation and included the Coriolis force, but we omit this here.
Thus, we have
\EQ
{\partial\meanFF\over\partial t}=
-\overline{u_i u_j}\,\partial\meanC
\underbrace
{-\overline{u_i u_j\partial_j c}
-\overline{c u_j\partial_j u_i}
-\overline{c \,\partial_i p}}
_{\mbox{triple correlations}\,\equiv\TT}.
\EN
This equation shows the important point that, at least in the steady state,
the triple correlations are {\it never} negligible.
Instead, they are actually comparable to the
quadratic correlation term on the right hand side.
(We remark that the first two terms in $\TT$ cancel if the volume
averages are used to
integrate by parts, but the third term still remains.)
The closure hypothesis used in MTA states that
\EQ
\TT=-{\meanFF\over\tau}\quad\mbox{(MTA closure hypothesis)}.
\EN
Inserting this expression and moving this $\meanFF$ to the left hand
side, we have
\EQ
\meanF_i=-\tau\overline{u_i u_j}\,\partial\meanC
-\tau{\partial\meanFF\over\partial t}.
\EN
The extra time derivative on the right hand side corresponds to
the Faraday displacement current in electrodynamics.
In the present case, this term can be neglected if the diffusion
speed is less than the turbulent rms velocity.
In some sense this is of course never the case, because with
ordinary Fickian diffusion (time derivative neglected) the diffusion
process is described by an elliptic equation which has infinite signal
propagation speed, and is hence violating causality.

In the following section we discuss recent work by Brandenburg et al.\
(2004) who showed that the presence of this displacement term in a
non-Fickian version of the diffusion equation is indeed justified.
Before turning to this aspect, let us contrast the MTA closure with
the FOSA closure assumption.
Here one is instead dealing with an integral equation,
\EQ
\meanF_i=-\int\overline{u_i(t)u_j(t')}\,\partial\meanC(t')\,\dd t'
+\mbox{triple correlations},
\label{FOSAresult}
\EN
where the triple correlations are neglected
[unless higher order terms are included; see Nicklaus \& Stix (1988)
or Carvalho (1992)].

In this particular case the two approaches become equivalent if
one assumes that the two-times correlation function $\overline{u_i(t)u_j(t')}$
is proportional to $\overline{u_iu_j}\exp[-(t-t')/\tau]$.
This can be shown by differentiating \Eq{FOSAresult}.
Thus, the causality problem in the Fickian diffusion approximation
stems really only from the commonly used approximation that the two-times
correlation function can be approximated by a delta function.
If the extra time derivative is not neglected, the diffusion equation
becomes a damped wave equation,
\begin{equation}
{\partial^2\meanC\over\partial t^2}
+{1\over\tau}{\partial\meanC\over\partial t}
=\onethird u_{\rm rms}^2\nabla^2\meanC,
\label{nonFickian_evol}
\end{equation}
where the wave speed is $u_{\rm rms}/\sqrt{3}$.
Note also that, after multiplication with $\tau$, the coefficient on the
right hand side becomes $\onethird\tau u_{\rm rms}^2\equiv\kappa_{\rm t}$,
and the second time derivative on the left hand side becomes unimportant
in the limit $\tau\to0$, or when the physical time scales are long compared
with $\tau$.
In that case we have simply
\begin{equation}
{\partial\meanC\over\partial t}
=\kappa_{\rm t}\nabla^2\meanC.
\label{elliptic}
\end{equation}
In the following we discuss a case where \Eq{elliptic} is clearly
insufficient and where the full time dependence has to be retained.

\subsection{Turbulent displacement flux and value of $\tau$}
\label{Sinitialflux}

\begin{figure}[t!]\begin{center}
\includegraphics[width=\columnwidth]{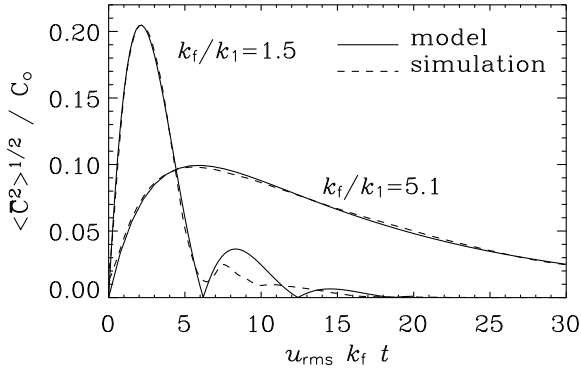}
\end{center}\caption[]{
Passive scalar amplitude, $\bra{\meanC^2}^{1/2}$,
versus time (normalized by $u_{\rm rms}k_{\rm f}$)
for two different values of $k_{\rm f}/k_1$.
The simulations have $256^3$ meshpoints.
The results are compared with solutions to the
non-Fickian diffusion model.
[Adapted from Brandenburg et al.\ (2004).]
}\label{plncc_comp}\end{figure}

A particularly obvious way of demonstrating the presence of the second
time derivative is by considering a numerical experiment where $\meanC=0$
initially.
\EEq{elliptic} would predict that then $\meanC=0$ at all times.
But, according to the alternative formulation \eq{nonFickian_evol},
this need not be true if initially $\partial\meanC/\partial t\neq0$.
In practice, this can be achieved by arranging the initial fluctuations
of $c$ such that they correlate with $u_z$.
Of course, such highly correlated arrangement will soon disappear and
hence there will be no turbulent flux in the long time limit.
Nevertheless, at early times, $\bra{\meanC^2}^{1/2}$ (an easily accessible
measure of the passive scalar amplitude) rises from zero to a finite
value; see \Fig{plncc_comp}.

\begin{figure}[t!]\begin{center}
\includegraphics[width=\columnwidth]{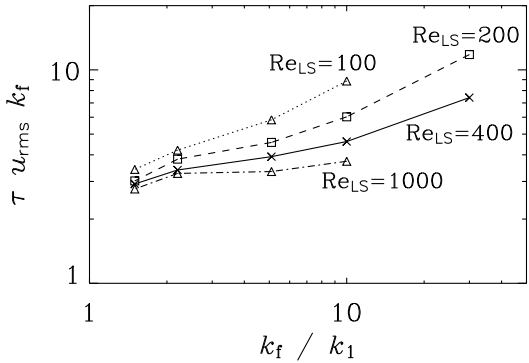}
\end{center}\caption[]{
Strouhal number as a function of $k_{\rm f}/k_1$ for different
values of $\mbox{Re}_{\rm LS}$, i.e.\ the large scale Reynolds number.
The resolution varies between $64^3$ meshpoints ($\mbox{Re}_{\rm LS}=100$)
and $512^3$ meshpoints ($\mbox{Re}_{\rm LS}=1000$).
}\label{pstrouhal_all_col}\end{figure}

Closer inspection of \Fig{plncc_comp} reveals that, when the wavenumber
of the forcing is sufficiently small (i.e.\ the size of the turbulent
eddies is comparable to the box size), $\bra{\meanC^2}^{1/2}$ approaches
zero in an oscillatory fashion.
This remarkable result can only be explained by the presence of
the second time derivative term giving rise to wave-like behavior.
This shows that the presence of the new term is actually justified.
Comparison with model calculations shows that the non-dimensional
measure of $\tau$, $\mbox{St}\equiv\tau u_{\rm rms} k_{\rm f}$,
must be around 3.
(In mean-field theory this number is usually called the Strouhal number.)
This rules out the validity of the quasilinear (first order smoothing)
approximation which would only be valid for $\mbox{St}\to0$.

Next, we consider an experiment to establish directly the value of St.
We do this by imposing a passive scalar gradient, which leads to a steady
state, and measuring the resulting turbulent passive scalar flux.
By comparing double and triple moments we can measure St quite accurately
without invoking a fitting procedure as in the previous experiment.
The result is shown in \Fig{pstrouhal_all_col} and it confirms that
$\mbox{St}\approx3$ in the limit of small forcing wavenumber, $k_{\rm f}$.
The details can be found in Brandenburg et al.\ (2004).

\subsection{Significance for the magnetic case}

In the hydromagnetic case one has
$\meanEMF=\overline{\uu\times\dot\bb}+\overline{\dot\uu\times\bb}$.
Here the $\dot\uu$ term is particularly important and usually not
obtained with FOSA.
Focusing on the term $\meanB_l b_{i,l}$ on the right hand side of the
$\dot u_i$ equation, we get
\EQ
\rho_0(\overline{\dot\uu\times\bb})_i
=\epsilon_{ijk}\overline{b_kb_{j,p}}\;\meanB_p+...
\stackrel{\mbox{iso}}{=}\;
\onethird\overline{\jj\cdot\bb}\;\meanB_i+...,
\EN
where the symbol $\;\stackrel{\mbox{iso}}{=}\;$ indicates isotropization
(which is really only done for simplicity and should be avoided
when it becomes important), and dots indicate the presence of further
terms which here only lead to triple correlation terms.
This $\overline{\jj\cdot\bb}$ term provides an important correction
to the usual $\alpha$ effect that results from
\EQ
(\overline{\uu\times\dot\bb})_i
=-\epsilon_{ijk}\overline{u_ku_{j,p}}\;\meanB_p+...
\!\stackrel{\mbox{iso}}{=}\!
-\onethird\overline{\oo\cdot\uu}\;\meanB_i+...
\EN
It is the $\overline{\jj\cdot\bb}$ current helicity term whose evolution,
under the assumption of scale separation, can be described in terms of
magnetic helicity conservation.
Since the magnetic helicity is conserved in the large magnetic Reynolds
number limit, the saturation time of a nonlinear dynamo can be very long.
This is now described in great detail in several recent reviews
(Brandenburg 2003, Brandenburg \& Subramanian 2004).
We emphasize however that the $\overline{\jj\cdot\bb}$ contribution is
quite decisive in describing correctly the slow saturation phase of any
nonlinear helical large scale dynamo in a periodic box
(Brandenburg 2001, Mininni et al.\ 2003).

\section{Problem III: $\alpha$ quenching}

\begin{figure}[b!]
\centering\includegraphics[width=\columnwidth]{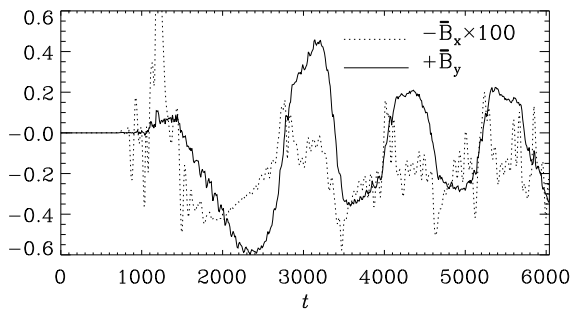}\caption{
Evolution of $\overline{B}_x$ and $\overline{B}_y$ at $x=-\pi$ and
$z=0$. Note that $\overline{B}_x$ has been scaled by a factor $-100$.
[Adapted from Brandenburg et al.\ (2001).]
}\label{Fpbutter_line}\end{figure}

\subsection{Non-universality of catastrophic quenching}

As shown in Brandenburg (2001), the slow saturation behavior of closed
box dynamos can be reasonably well be described in a mean field model
using catastrophic quenching, i.e.\ $\alpha(\meanBB)=\alpha_0 Q(\meanBB)$
and $\eta_{\rm t}(\meanBB)=\eta_{\rm t0} Q(\meanBB)$, where
$Q(\meanBB)=(1+R_{\rm m}\meanBB^2\!/B_{\rm eq}^2)^{-1}$.
As discussed in detail by Blackman \& Brandenburg (2002), the reason for
the agreement with the simulations is due to the `force-free degeneracy'
of the $\alpha^2$ dynamo in a periodic box, because then $\meanBB$ and
$\meanJJ$ are then everywhere parallel to each other.
This degeneracy is lifted for $\alpha\Omega$ dynamos, i.e.\ in the
presence of shear.

\begin{figure}[t!]
\centering\includegraphics[width=\columnwidth]{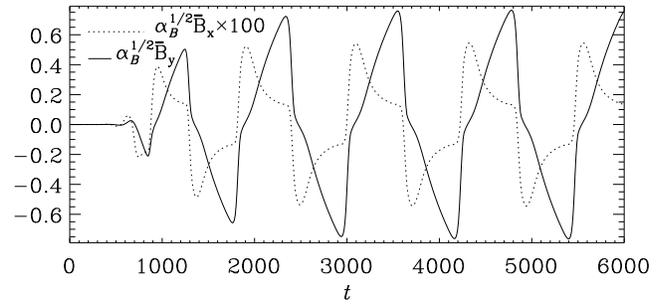}\caption{
Evolution of $\alpha_B^{1/2}\overline{B}_x$ and
$\alpha_B^{1/2}\overline{B}_y$ in the
one-dimensional mean-field model with a dynamo number ${\cal D}=10$,
a kinematic growth rate $\lambda=0.015$ (which determines $\alpha$)
and a microscopic magnetic diffusivity $\eta=5\times10^{-4}$.
Note that $\overline{B}_x$ has been scaled by a factor $100$.
(In this case the shear $S>0$, so we have plotted $+B_x$,
and not $-B_x$ as we did in \Fig{Fpbutter_line} where $S<0$.)
[Adapted from Brandenburg et al.\ (2001).]
}\label{Fptime}\end{figure}

\begin{figure}[t!]
\centering\includegraphics[width=\columnwidth]{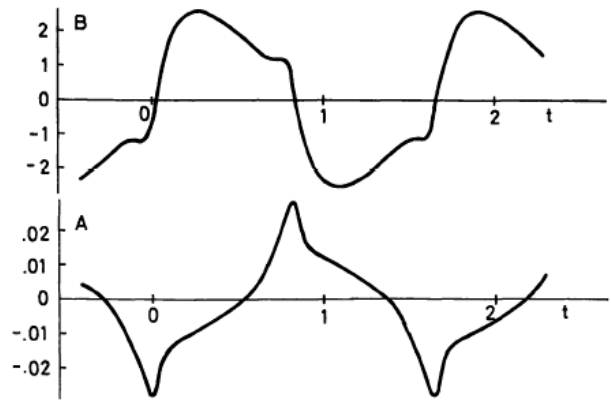}\caption{
Evolution of toroidal and poloidal fields, $B$ and $A$, respectively,
from the paper by Stix (1972).
Note the similarity in the evolution of his model and that in
the direct simulation shown in \Fig{Fpbutter_line}, with a more spiky
behavior of the poloidal field compared to the toroidal field.
[Adapted from Stix (1972).]
}\label{stix72b}\end{figure}

In \Fig{Fpbutter_line} we show the evolution of poloidal and toroidal
fields, $B_x$ and $B_y$, at one point in the simulation of Brandenburg
et al.\ (2001), which has shear.
Note the systematic phase shift and a well-defined amplitude ratio
between $B_x$ and $B_y$.
Note also that the dynamo wave is markedly non-harmonic.
These are clear properties that can be compared with mean-field model
calculations; see \Fig{Fptime}.
A similar type of non-harmonic temporal behavior has been found in
the first-ever nonlinear simulation of a mean field dynamo by Stix (1972);
see also \Fig{stix72b}.
An important difference between the two models is that for the results
shown in \Fig{Fptime} both $\alpha$ and $\eta_{\rm t}$ where quenched,
whereas in the model shown in \Fig{stix72b} $\eta_{\rm t}$ was kept
constant and $\alpha$ was quenched in a step function-like fashion.
Since the results from these two very different models are
similar, the temporal behavior alone cannot really be used to discriminate
one model over another.

\subsection{Dynamical $\alpha$ quenching}

The $\alpha$ effect formalism provides so far the only workable
mathematical framework for describing the large scale dynamo action seen
in simulations of helically forced turbulence.
The governing equation for the mean magnetic field is
\EQ
{\partial\meanBB\over\partial t}=\nab\times\left(
\meanUU\times\meanBB+\meanEMF-\eta\mu_0\meanJJ\right),
\EN
where $\meanEMF=\overline{\uu\times\bb}$ is the electromotive
force resulting from the $\uu\times\bb$ nonlinearity in the
averaged Ohm's law.
MTA can be used to derive an expression for $\meanEMF$ in terms
of the mean field.
For slow rotation one finds (R\"adler et al.\ 2003, see also
review by Brandenburg \& Subramanian 2004)
\EQ
\meanemf_i = \alpha_{ij} \meanB_j - \eta_{ij} \meanJ_j
+ (\ggamma\times\meanBB+\ddelta\times\meanJJ)_i
+ \kappa_{ijk} \meanB_{j,k},
\label{emf_rot}
\EN
Under isotropic conditions, or as a means of simplification,
one often just writes
\EQ
\meanEMF=\alpha\meanBB-\eta_{\rm t}\meanJJ
\label{emf_iso}
\EN
In any case, and this comes as a rather recent realization
(Field \& Blackman 2002, Blackman \& Brandenburg 2002, Subramanian 2002),
there will be an extra magnetic contribution to the $\alpha$ effect
$\alpha\to\alpha=\alpha_{\rm K}+\alpha_{\rm M}$.
In the isotropic case, $\alpha_{\rm K}=-\onethird\tau\overline{\oo\cdot\uu}$
and $\alpha_{\rm M}=\onethird\tau\overline{\jj\cdot\bb}/\rho$.
This leads to (Kleeorin \& Ruzmaikin 1982; see also
Zeldovich et al.\ 1983; and Kleeorin et al.\ 1995)
\EQ
{\dd\alpha_{\rm M}\over\dd t}=-2\eta_{\rm t0}k_{\rm f}^2\left(
{\bra{\meanEMF\cdot\meanBB}\over B_{\rm eq}^2}
+{\alpha_{\rm M}\over\tilde{R}_{\rm m}}\right),
\label{dynquench}
\EN
where we have defined the magnetic Reynolds number as
\EQ
\tilde{R}_{\rm m}=\eta_{\rm t0}/\eta,
\label{Rmdef}
\EN
where $\eta_{\rm t0}=\onethird\tau\bra{\uu^2}$ is the kinematic
value of the turbulent magnetic diffusivity and
$B_{\rm eq}^2=\mu_0\rho_0\bra{\uu^2}$ is the equipartition field strength.
\EEq{dynquench} agrees with the corresponding equation
in Kleeorin et al.\ (1995) if their
characteristic length scale of the turbulent motions at the
surface, $l_{\rm s}$, is identified with
$2\pi/k_{\rm f}$ and if their parameter $\mu$ is identified
with $8\pi^2\eta_{\rm t0}^2/(\bra{\uu^2}l_{\rm s}^2)$.
The definition of $\tilde{R}_{\rm m}$ may not be very practical if
$\eta_{\rm t0}$ is not known, but comparison with simulations
(Blackman \& Brandenburg 2002) suggests that
$\tilde{R}_{\rm m}\approx0.3R_{\rm m}$, where
$R_{\rm m}=u_{\rm rms}/(\eta k_{\rm f})$ is a more practical definition
suitable for simulations of forced turbulence.

The basic idea is that magnetic helicity conservation must be obeyed,
but the presence of an $\alpha$ effect leads to magnetic helicity of
the mean field which has to be balanced by magnetic helicity of
the fluctuating field.
This magnetic helicity of the fluctuating (small scale) field must
be of opposite sign to that of the mean (large scale) field
(e.g.\ Blackman \& Brandenburg 2003).
In the following we refer to \Eq{dynquench} as the the dynamical
$\alpha$ quenching equation (Blackman \& Brandenburg 2002).
Assuming that the large scale magnetic field has reached a steady state,
and solving this equation for $\alpha$, yields (Kleeorin \& Ruzmaikin 1982,
Gruzinov \& Diamond 1994)
\EQ
\alpha={\alpha_{\rm K}
+\eta_{\rm t} R_{\rm m}\bra{\meanJJ\cdot\meanBB}/B_{\rm eq}^2
\over1+R_{\rm m}\bra{\meanBB^2}/B_{\rm eq}^2}
\quad\mbox{(for $\dd\alpha/\dd t=0$)}.
\label{AlphaStationary}
\EN
Note that for the numerical experiments with an imposed large scale field
over the scale of the box (Cattaneo \& Hughes 1996), where $\meanBB$
is spatially uniform and therefore $\meanJJ=0$, one recovers the
`catastrophic' quenching formula,
\EQ
\alpha={\alpha_{\rm K}\over1+R_{\rm m}\bra{\meanBB^2}/B_{\rm eq}^2}
\quad\mbox{(for $\meanJJ=0$)},
\EN
which implies that $\alpha$ becomes quenched when
$\bra{\meanBB^2}/B_{\rm eq}^2=R_{\rm m}^{-1}\approx10^{-8}$
for the sun, and for even smaller fields in the case of galaxies.

Obviously, the assumption of a steady state is generally not permitted.
Especially in the case of closed boxes that are so popular for simulation
purposes, there is no other way to get rid of magnetic helicity than
via microscopic resistivity.
There is now a significant body of literature (see Brandenburg 2001
for a comprehensive coverage of the simulation results and
Brandenburg 2003 for a recent review).
To allow for faster time scales, and this was realized first by
Blackman \& Field (2000a,b) and Kleeorin et al.\ (2000), there might
be hope to achieve this by allowing helicity flux to escape through
the boundaries of the domain.
Two types of results will be discussed in the next section.

\section{Open boundaries and shear: the solution?} 
\label{alphaOPEN} 

\begin{figure}[t!]
\centering\includegraphics[width=\columnwidth]{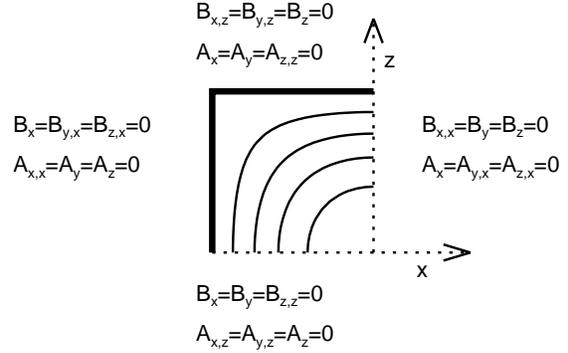}\caption{
Differential rotation in our cartesian model, with the equator
being at the bottom, the surface to the right, the bottom of the
convection zone to the left and mid-latitudes at the top.
[Adapted from Brandenburg \& Sandin (2004).]
}\label{sketch1}\end{figure}

In a recent paper, Brandenburg \& Sandin (2004) have carried out a
range of simulations for different values of the magnetic Reynolds
number, $R_{\rm m}=u_{\rm rms}/(\eta k_{\rm f})$, for both open and
closed boundary conditions using the geometry depicted in
\Fig{sketch1}.  In the following we discuss first some results for
$\alpha$ quenching and an interpretation of the results in terms of
the current helicity flux.  Next we turn to direct simulations of the
dynamo without an imposed field.

\subsection{Results for $\alpha$ quenching}
\label{ResultsAlphaQuenching} 

In order to measure $\alpha$, a uniform magnetic field,
$\BB_0=\const$, is imposed, and the magnetic field is now written as
$\BB=\BB_0+\nab\times\AAA$.  Brandenburg \& Sandin (2004) determined
$\alpha$ by measuring the turbulent electromotive force, and hence
$\alpha=\bra{\emf}\cdot\BB_0/B_0^2$.  Similar investigations have been
done before both for forced turbulence (Cattaneo \& Hughes 1996,
Brandenburg 2001) and for convective turbulence (Brandenburg et
al.\ 1990, Ossendrijver et al.\ 2001).

As expected, $\alpha$ is negative when the helicity of the forcing is
positive, and $\alpha$ changes sign when the helicity of the forcing
changes sign.
The magnitudes of $\alpha$ are however different in the two cases:
$|\alpha|$ is larger when the helicity of the forcing is negative.
In the sun, this corresponds to the sign of helicity in the northern
hemisphere in the upper parts of the convection zone.
This is here the relevant case, because the differential rotation
pattern of the present model also corresponds to the northern hemisphere.

\begin{figure}[t!]
\centering\includegraphics[width=.9\columnwidth]{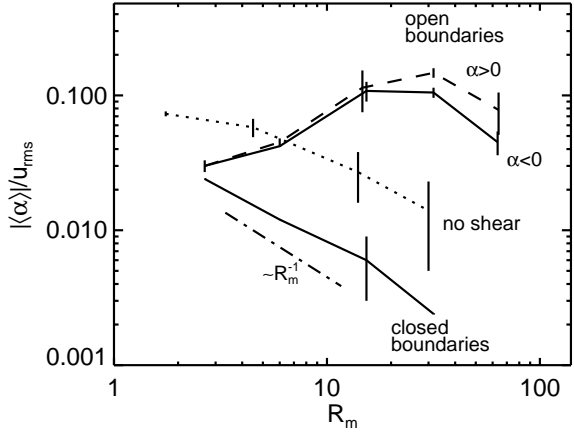}\caption{
Dependence of $|\bra{\alpha}|/u_{\rm rms}$ on $R_{\rm m}$
for open and closed boundaries.
The case with open boundaries and negative helicity is shown as a dashed line.
Note that for $R_{\rm m}\approx30$ the $\alpha$ effect
is about 30 times smaller when the boundaries are closed.
The dotted line gives the result with open boundaries but no shear.
The vertical lines indicate the range obtained by calculating
$\alpha$ using only the first and second half of the time interval.
[Adapted from Brandenburg \& Sandin (2004).]
}\label{palp_sum}\end{figure}

There is a striking difference between the cases with open and closed
boundaries which becomes particularly clear when comparing the
averaged values of $\alpha$ for different magnetic Reynolds numbers;
see \Fig{palp_sum}.  With closed boundaries $\alpha$ tends to zero
like $R_{\rm m}^{-1}$, while with open boundaries $\alpha$ shows no
such decline.  There is also a clear difference between the cases with
and without shear together with open boundaries in both cases.  In the
absence of shear (dotted line in \Fig{palp_sum}) $\alpha$ declines
with increasing $R_{\rm m}$, even though for small values of $R_{\rm
  m}$ it is larger than with shear.  The difference between open and
closed boundaries will now be discussed in terms of a current helicity
flux through the two open boundaries of the domain.

\subsection{Interpretation in terms of current helicity flux}

It is suggestive to interpret the above results in terms of the
dynamical $\alpha$ quenching model.
However, the quenching equation has to be generalized to take the divergence
of the flux into account.
In order to avoid problems with the gauge, it is advantageous to
work directly with $\overline{\jj\cdot\bb}$ instead of
$\overline{\aaaa\cdot\bb}$.
Using the evolution equation, $\partial\bb/\partial t=-\nab\times\ee$,
for the fluctuating magnetic field, where $\ee=\EE-\meanEE$ is the
small scale electric field and $\meanEE=\eta\mu_0\meanJJ-\meanEMF$ the
mean electric field, one can derive the equation
\begin{equation}
{\partial\over\partial t}\overline{\jj\cdot\bb}
=-2\,\overline{\ee\cdot\cc}-\nab\cdot\meanFF_C^{\rm SS},
\label{jc_evolution}
\end{equation}
where
\begin{equation}
\meanFF_C^{\rm SS}=\overline{2\ee\times\jj}
+\overline{(\nab\times\ee)\times\bb}/\mu_0,
\label{meanFFc}
\end{equation}
is the current helicity flux from the small scale (SS) field, and
$\cc=\nab\times\jj$ the curl of the small scale current density,
$\jj=\JJ-\meanJJ$.
In the isotropic case, 
$\overline{\ee\cdot\cc}\approx k_{\rm f}^2\overline{\ee\cdot\bb}$, where
$k_{\rm f}$ is the typical wavenumber of the fluctuations,
here assumed to be the forcing wavenumber.

Making use of the adiabatic approximation one arrives at the algebraic
steady state quenching formula ($\partial\alpha/\partial t=0$)
\begin{equation}
\alpha={\alpha_{\rm K}
+R_{\rm m}\left(\eta_{\rm t}\mu_0\meanJJ\cdot\meanBB
-\half k_{\rm f}^{-2}\nab\cdot\mu_0\meanFF_C^{\rm SS}\right)/B_{\rm eq}^2
\over1+R_{\rm m}\meanBB^2/B_{\rm eq}^2}.
\label{AlphaStationaryFlux}
\end{equation}
In the absence of a mean current, e.g.\ if the mean field is defined
as an average over the whole box, then $\meanBB\equiv\BB_0=\const$,
and $\meanJJ=0$, so \Eq{AlphaStationaryFlux} reduces to
\begin{equation}
\alpha={\alpha_{\rm K}
-\half k_{\rm f}^{-2}R_{\rm m}\nab\cdot\mu_0\meanFF_C^{\rm SS}/B_{\rm eq}^2
\over1+R_{\rm m}\BB_0^2/B_{\rm eq}^2}.
\label{AlphaStationaryFlux_noJ}
\end{equation}
This expression applies to the present case, because we consider
only the statistically steady state and we also define the mean field
as a volume average.

In the simulations, the current helicity flux is found to be independent
of the magnetic Reynolds number.
This explains why the $\alpha$ effect no longer shows the catastrophic
$R_{\rm m}^{-1}$ dependence (see \Fig{palp_sum}).
In principle it is even conceivable that with $\alpha_{\rm K}=0$
a current helicity flux can be generated, for example by shear,
and that this flux divergence could drive a dynamo, as was suggested
by Vishniac \& Cho (2001).
It is clear, however, that for finite values of $R_{\rm m}$ this would
be a non-kinematic effect requiring the presence of an already finite
field (at least of the order of $B_{\rm eq}/R_{\rm m}^{1/2}$).
This is because of the $1+R_{\rm m}\BB_0^2/B_{\rm eq}^2$ term in the
denominator of \Eq{AlphaStationaryFlux_noJ}.
At the moment we cannot say whether this is perhaps the effect leading
to the nonhelically forced turbulent dynamo discussed in \Sec{OpenSurface},
or whether it is perhaps the $\ddelta\times\meanJJ$ or shear-current
effect that was also mentioned in that section.

\subsection{Dynamos with open surfaces and shear}
\label{OpenSurface}

The presence of an outer surface is in many respects similar to the
presence of an equator.
In both cases one expects magnetic and current helicity fluxes via
the divergence term.
A particularly instructive system is helical turbulence in an
infinitely extended horizontal slab with stress-free boundary
conditions and a vertical field condition, i.e.\
\EQ
u_{x,z}=u_{y,z}=u_{z}=B_x=B_y=0.
\EN
Without shear,
such simulations have been performed by Brandenburg \& Dobler (2001) who
found that a mean magnetic field is generated, similar to the case with
periodic boundary conditions, but that the energy of the mean magnetic
field, $\bra{\meanBB^2}$, decreases with magnetic Reynolds number.
Nevertheless, the energy of the total magnetic field, $\bra{\BB^2}$,
does not decrease with increasing magnetic Reynolds number.
Although they found that $\bra{\meanBB^2}$ decreases only like $R_{\rm
m}^{-1/2}$, new simulations confirm that a proper scaling regime has
not yet been reached and that the current data may well be compatible
with an $R_{\rm m}^{-1}$ dependence.

\begin{figure}[t!]
\centering\includegraphics[width=\columnwidth]{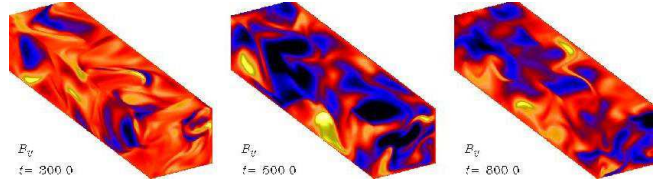}\caption{
Visualization of the toroidal magnetic field during three different times
during the growth and saturation for the run without kinetic helicity.
}\label{diffrot_small}\end{figure}

Clearly, an asymptotic decrease of the mean magnetic field must mean
that the small scale dynamo does not work with such boundary conditions.
Thus, the anticipated advantages of open boundary conditions are not
borne out by this type of simulations.
In the presence of shear the results are very different.
We now use the same setup as in \Sec{ResultsAlphaQuenching}.
The size of the computational domain is $\half\pi\times2\pi\times\half\pi$
and the numerical resolution is $128\times512\times128$ meshpoints.
The magnetic Reynolds number based on the forcing wavenumber and the
turbulent flow is around 80 and shear flow velocity exceeds the rms
turbulent velocity by a factor of about 5.
We have carried out experiments with no helicity in the forcing
(labeled as $\alpha=0$), as well as positive and negative helicity
in the forcing (labeled $\alpha<0$ and $\alpha>0$, respectively);
see \Fig{diffrot_small} for a visualization of the run without
kinetic helicity.
We emphasize that no explicit $\alpha$ effect has been invoked.
The labeling just reflects the fact that, in isotropic turbulence, negative
kinetic helicity (as in the northern hemisphere of a star or the upper
disc plane in galaxies) leads
to a positive $\alpha$ effect, and vice versa.

\begin{figure}[t!]
\centering\includegraphics[width=.9\columnwidth]{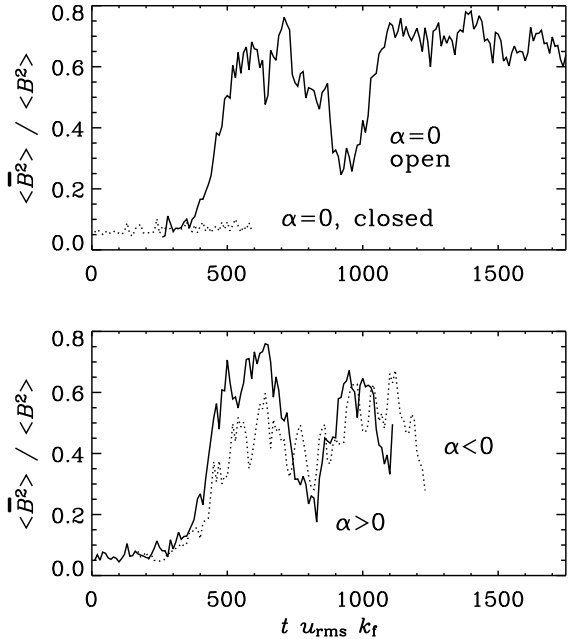}\caption{
Saturation behavior of the ratio $q=\bra{\meanBB^2}/\bra{\BB^2}$
for runs with different kinetic helicity of the flow.
Solid line: zero helicity,
dotted line: positive helicity (opposite to the sun)
dashed line: negative helicity (as in the sun).
The line at denoted by ``$\alpha=0$, closed'' refers to a case
where the normal field condition on the equator and the surface
has been replaced by a perfect conductor condition.
}\label{pmean_comp}\end{figure}

\begin{figure}[t!]
\centering\includegraphics[width=.9\columnwidth]{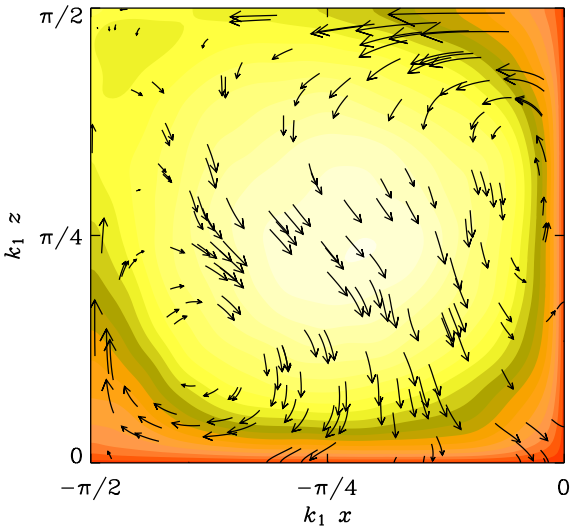}\caption{
Gray/color scale representation of the azimuthally and time averaged mean
azimuthal field $\meanBB$ together with vectors in the meridional plane.
In this run the turbulence is nonhelically forced ($\alpha=0$).
}\label{pbm}\end{figure}

\begin{figure}[t!]
\centering\includegraphics[width=.9\columnwidth]{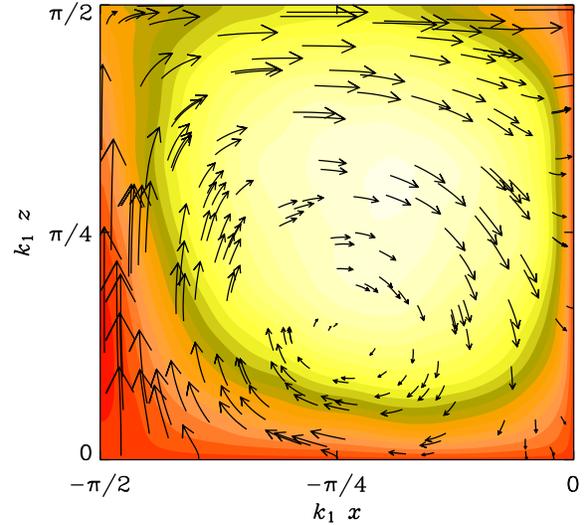}\caption{
Same as \Fig{pbm}, but for $\alpha>0$, so
the turbulence is forced with negative helicity.
Note the clockwise sense of the poloidal field.
Together with the positive toroidal field (indicated by light shades) this
corresponds to positive magnetic helicity, consistent with $\alpha>0$.
}\label{pbm_apos}\end{figure}

\begin{figure}[t!]
\centering\includegraphics[width=.9\columnwidth]{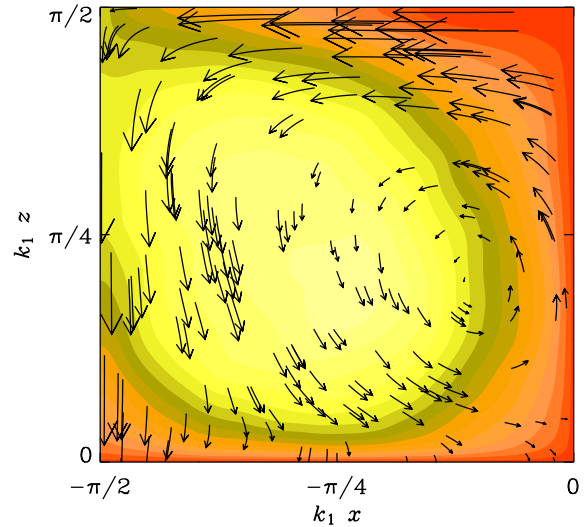}\caption{
Same as \Fig{pbm}, but for $\alpha<0$, so
the turbulence is forced with positive helicity.
Note the anti-clockwise sense of the poloidal field,
corresponding to negative magnetic helicity, consistent with $\alpha>0$.
}\label{pbm_aneg}\end{figure}

\begin{figure}[t!]
\centering\includegraphics[width=.9\columnwidth]{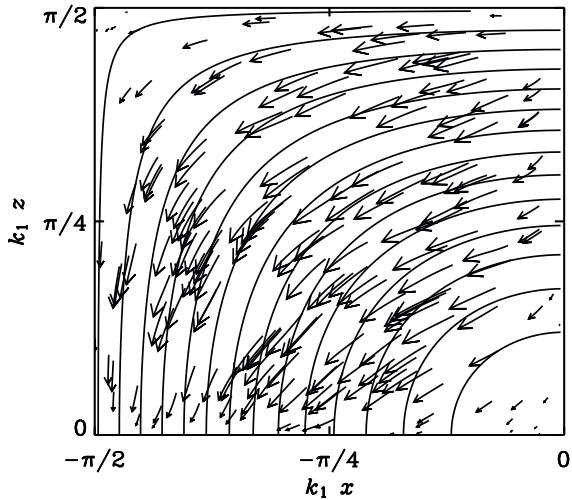}\caption{
Vectors of $\phi_{iyy}$ together with contours of $\meanU_y$ which
also coincide with the streamlines of the mean vorticity field $\meanWW$.
Note the close agreement between $\phi_{iyy}$ and $\meanWW$.
The orientation of the vectors indicates that negative current helicity
leaves the system at the outer surface ($x=0$).
}\label{p2vishcho}\end{figure}

We characterize the relative strength of the mean field by the
ratio $q=\bra{\meanBB^2}/\bra{\BB^2}$, where overbars denote an average
in the toroidal ($y$) direction; see \Fig{pmean_comp}.
A strong mean field is generated in all cases, unless a perfect conductor
boundary condition (``closed'') is adopted on the outer surface and on
the equator.
The mean field appears to be statistically stationary in all cases,
i.e.\ there is no indication of migration in the meridional plane.
A time average of the mean field is shown in \Fig{pbm}.

There are two surprising results emerging from this work.
First, in the presence of shear rather strong mean fields can be
generated, where up to 70\% of the energy can be in the mean field;
see \Fig{pmean_comp}.
Second, even without any kinetic helicity in the flow there is strong
large scale field generation.
However, this cannot be an $\alpha\Omega$ dynamo in the usual sense.
One possibility is the $\ddelta\times\JJ$ effect, which emerged
originally in the presence of the Coriolis force; see R\"adler (1969)
and Krause \& R\"adler (1980).
In the present case with no Coriolis force, however, a $\ddelta\times\JJ$
effect is possible even in the presence of shear alone, because
the vorticity associated with the shear contributes directly to
$\ddelta\propto\meanWW=\nab\times\meanUU$ (Rogachevskii \& Kleeorin 2003, 2004).

A rather likely candidate for the current helicity flux is the so-called
Vishniac \& Cho (2001) flux.
A systematic derivation using MTA (Subramanian \& Brandenburg 2004) gave
\EQ
\meanF^{\rm SS}_{C\,i}=\phi_{ijk}\meanB_j\meanB_k,
\EN
where
$\phi_{ijk}$ is a new turbulent transport tensor with
\EQ
\phi_{ijk}
=-4\tau\overline{\omega_k \nabla_j u_i}.
\label{vishflux}
\EN
Obviously, only components of $\phi_{ijk}$ that are symmetric in
$j$ and $k$ enter the flux $\meanFF_C^{\rm SS}$.
Therefore we consider in the following
$\phi_{ijk}^{\rm S}=\half(\phi_{ijk}+\phi_{ikj})$.
In the present case where the toroidal field ($y$ direction)
is strongest, the $\phi_{iyy}$ components (where $i=x$ and $z$)
are expected to be most important.
The time average of this component is shown in \Fig{p2vishcho}.
Here we have only considered the nonhelical case, but the helical
cases are indistinguishable within error margins.
All the 6 independent components of $\phi_{xij}^{\rm S}$ and
$\phi_{zjk}^{\rm S}$ are given by
\EQ
\phi_{xjk}^{\rm S}=\pmatrix{
 -0.03 & +0.04 & -0.12 \cr
 +0.04 & -0.23 & -0.36 \cr
 -0.12 & -0.36 & +0.25
},
\EN
\EQ
\phi_{zjk}^{\rm S}=\pmatrix{
 +0.24 & +0.32 & -0.13 \cr
 +0.32 & -0.19 & -0.05 \cr
 -0.13 & -0.05 & -0.04
}.
\EN
Note that the trace of both tensors is small.
Thus there is no hope that this tensor could possibly be isotropic.
This is because isotropization, i.e.\ $\onethird\delta_{jk}\phi_{ijk}$,
would then give something close to zero.
Nevertheless, one may hope that some useful parameterization of
$\phi_{ijk}^{\rm S}$, that can be used in mean field calculations,
will soon be available.

In conclusion,
there is evidence that the strong dynamo action seen in the
simulations is only possible due to the combined presence of open
boundaries and shear.

\section{Conclusions}

In the present work we have discussed three aspects that are strongly
connected with the applicability of mean field dynamo theory.
The first concerns turbulent diffusion, i.e.\ the ability of the flow to
mix a large scale magnetic field such that its energy can be converted
into energy at smaller scales.
In order that microscopic diffusion can eventually be thermalized,
unimpeded by the Lorentz force, the field has to be weak enough
at the scale where diffusion takes place.
Whether this is indeed the case can in principle be seen from the simulations.
Since in the simulations the magnetic and fluid Reynolds numbers are much
smaller than in reality, it would be important to show that the
spectra begin to converge for large magnetic and fluid Reynolds numbers.
At the moment this is simply not yet the case, even at a resolution of
$1024^3$ meshpoints.
However, by a combination between direct and large eddy simulations
some preliminary insight can be gained.
The suggestions are that magnetic and kinetic energy spectra approach
spectral equipartition in the deep inertial range, and that the slope
would be comparable to the Kolmogorov $k^{-5/3}$ slope.

The other aspect discussed in this paper concerns the calculation of
turbulent transport coefficients such as the $\alpha$ effect.
It has been particularly disturbing that much of the theory in this field
was based on the first order smoothing approximation which clearly should
break down under physically interesting conditions.
The reason for its apparent success is now becoming clear and are
connected with the close similarity between the first order smoothing
approximation and the minimal tau approximation where the triple
correlations are no longer omitted.
The reason for the close similarity between the two approaches is
mainly connected with the fact that all the interesting physics enters
via the quadratic correlations which can be worked out correctly using
linear theory.
The significance of the triple correlations is merely that they
determine the length of $\tau$.
In the framework of the first order smoothing approximation the value
of $\tau$ had to be obtained by dimensional arguments and was therefore
not well determined.

Next we have briefly addressed the long standing issue of $\alpha$
quenching and have pointed out the connection between dynamical quenching
and catastrophic quenching.
The possibility of catastrophic $\alpha$ quenching is closely linked
to magnetic helicity conservation.
The degree of quenching can therefore be alleviated by allowing for
magnetic or better current helicity fluxes.
Such fluxes do not just come on their own, so simply allowing for the
boundaries to be open is not sufficient.
One needs to drive a flux through the entire domain.
This seems to be accomplished by the Vishniac-Cho flux.
As already argued by Vishniac \& Cho (2001), and confirmed by
Arlt \& Brandenburg (2001), this flux can be driven in the presence
of shear.
Indeed, there are strong indications that the helicity flux
can alleviate catastrophic quenching at least by a factor of
about 30, and that it allows for dynamo action generating strong
magnetic fields, even in the absence of helicity in the forcing.
The details of this process are not entirely clear, but
possible candidates include the shear-current effect of
Rogachevskii \& Kleeorin (2003, 2004), and perhaps the helicity flux
itself (Vishniac \& Cho 2001) which is a nonlinear effect.

In the near future it should be possible to investigate the emergence
of current helicity fluxes from a dynamo simulation in more detail.
This would be particularly interesting in view of the many observations of
coronal mass ejections that are known to be associated with significant
losses of magnetic helicity and hence also of current helicity
(Berger \& Ruzmaikin 2000, DeVore 2000, Chae 2000, Low 2001,
D\'emoulin et al.\ 2002, Gibson et al.\ 2002).
In order to be able to model coronal mass ejections it should be
particularly important to relax the restrictions imposed by the vertical
field conditions employed in the simulations of Brandenburg \& Sandin (2004).
One possibility would be to include a simplified version
of a corona with enhanced temperature and hence decreased density,
giving rise to a low-beta plasma exterior where nearly force-free fields
can develop (Gudiksen \& Nordlund 2003).
Such a setup might allow detailed comparison with observations.

\acknowledgements

The Danish Center for Scientific Computing is acknowledged
for granting time on the Linux cluster in Odense (Horseshoe).


\end{document}